\documentclass[12pt]{revtex4}
\usepackage[dvips]{graphicx}
\usepackage{amssymb}
\usepackage{amsmath}
\usepackage{color}
\usepackage{float}

\newcommand {\apgt} {\ {\raise-.5ex\hbox{$\buildrel>\over\sim$}}\ }
\newcommand {\aplt} {\ {\raise-.5ex\hbox{$\buildrel<\over\sim$}}\ } 

\newcommand{\epso}{\varepsilon_0}
\newcommand{\dy}{\mbox{d}y}
\newcommand{\vp}{v_{\perp}}
\newcommand{\vpa}{v_{\|}}
\begin{document}
\begin{center}
  \title{Quasi-linear analysis of the extraordinary electron wave
    destabilized by runaway electrons}

  \author{G.I. Pokol$^{1}$, A. K\'om\'ar$^{1}$, A. Budai$^{1}$,
    A. Stahl$^{2}$, T. F\"ul\"op$^{2}$}

  \affiliation{
    \small $^{1}$ Department of Nuclear Techniques, Budapest University of Technology and Economics, Budapest, Hungary\\
%\small $^{2}$ California Institute of Technology, Pasadena, USA\\
\small $^{2}$ Department of Applied Physics,  Chalmers University of Technology,  G\"oteborg, Sweden\\
}

\maketitle

\end{center}
\begin{abstract}
  Runaway electrons with strongly anisotropic distributions present in
  post-disruption tokamak plasmas can destabilize the extraordinary
  electron (EXEL) wave.  The present work investigates the dynamics of
  the quasi-linear evolution of the EXEL instability for a range of
  different plasma parameters using a model runaway distribution
  function valid for highly relativistic runaway electron beams
  produced primarily by the avalanche process. Simulations show a
  rapid pitch-angle scattering of the runaway electrons in the high
  energy tail on the $100-1000\;\rm \mu s$ time scale. Due to the
  wave-particle interaction, a modification to the synchrotron
  radiation spectrum emitted by the runaway electron population is
  foreseen, exposing a possible experimental detection method for such
  an interaction.
\end{abstract}
\maketitle

\section{Introduction}

Disruptions in tokamaks can lead to the generation of a high-current
beam of highly energetic runaway electrons \cite{hender}, which poses
great challenges for the disruption mitigation system of future
tokamaks \cite{hollmann}. The runaway electron beam has a strongly
anisotropic velocity distribution and may destabilize high-frequency
electromagnetic and electrostatic waves through a resonant
interaction. Several high-frequency instabilities driven by runaway
electrons have been considered before, using various models for the
initial runaway distribution function
\cite{parail,breizman,komarpop2012,lai2013,fuloppop2006,pokolppcf2008}. In
particular, the linear stability and the quasi-linear analysis of the
whistler wave instability (WWI) has been investigated, and it was
concluded that whistler waves may be destabilized by an avalanching
runaway electron population \cite{fuloppop2006,pokolppcf2008}. The
main motivation for that work was to investigate the possible effect
of these waves on the runaway electron beam formation. If such an
instability would lead to scattering of the runaway electrons in
pitch-angle, resulting in higher synchrotron radiation
losses, a passive
mitigation mechanism limiting the detrimental effects of the runaway
electrons would be provided. However, it was concluded that for the low temperatures
characteristic of post-disruption plasmas, the collisional damping is
likely to suppress the WWI and the effect of the instability on runaway beam formation is therefore small. On the other hand, the WWI may provide a
diagnostic opportunity due to its sensitive dependence on the fast
electron distribution function and the plasma parameters.

Recently, it has been shown that runaways can also destabilize so-called extraordinary-electron (EXEL) waves at oblique propagation
angles \cite{komarvarenna}. Compared to the WWI, it was found that significantly fewer energetic electrons are needed to destabilize the EXEL wave, which is therefore likely to be the most unstable wave~\cite{komarvarenna}.
The aim of this work is to determine the
characteristics of the quasi-linear evolution of the EXEL instability
and quantify its effects on the runaway electron beam. We also
investigate the possibility of detecting signatures of the
wave-particle interaction in the experimental infrared synchrotron
emission data.

In large tokamak disruptions, where the principal source of runaway
electrons is the secondary avalanche process \cite{rosenputv}, an analytical distribution
function for the runaway electrons (in the
absence of wave-particle interaction) can be obtained
\cite{fuloppop2006}. This distribution function has been benchmarked
to the results of numerical simulations \cite{code} and has been used
in Ref.~\cite{pokolppcf2008} as an initial runaway distribution
function for the quasi-linear evolution of the WWI.
In the present work we adopt a similar approach, extending the treatment to the EXEL wave. 
 
 One possible method of inferring the characteristics of the runaway population is to study the synchrotron radiation emitted by the energetic electrons. By calculating the integrated emission from the complete electron population \cite{adamsyrup}, we show that the pitch-angle scattering of
highly energetic runaway electrons due to the interaction with the
EXEL wave causes a characteristic change in the synchrotron spectrum
that could be detected in experiments.

The structure of the paper is as follows.  In Sec.~\ref{sec:II} the
dispersion relation and the characteristics of the EXEL wave are
described. In Sec.~\ref{sec:III} we investigate the quasi-linear
evolution of the EXEL wave and its effect on the distribution of fast
electrons. Section~\ref{sec:IV} completes the analysis with a study of
the parametric dependencies of the process. The calculations of the
synchrotron spectrum of the affected distribution, presented in
Sec.~\ref{sec:V}, provide guidelines for possible experimental
detection of the instability. Finally, the results are discussed and
summarized in Sec.~\ref{sec:conclusions}.

\section{Excitation of the extraordinary electron wave}
\label{sec:II}
The characteristics of the EXEL wave can be derived from the wave
dispersion relation in a homogeneous, magnetized plasma approximation
\cite{stix}:
\begin{equation}
\left[ \left(\epsilon_{11} - \frac{k_\parallel^2 c^2}{\omega^2}\right) \left(\epsilon_{22} - \frac{k^2 c^2}{\omega^2}\right) + 
    \epsilon_{12}^2 \right] \left(\epsilon_{33} - \frac{k_\perp^2 c^2}{\omega^2}\right) - \frac{k_\parallel^2 k_\perp^2 c^4}{\omega^4} \left(\epsilon_{22} - \frac{k^2 c^2}{\omega^2}\right) = 0.
\label{eq:fulldisp}
\end{equation}
Note that, to describe the EXEL wave, the frequently used
electromagnetic approximation $\epsilon_{33} \gg (k c/ \omega)^2
\cos{\theta} \sin{\theta}$ has to be relaxed. Here $k$ is the wave
number, $k_\parallel$ and $k_\perp$ denote its components parallel and
perpendicular to the static magnetic field, respectively, and $\cos
\theta=k_\parallel/k$. $\omega$ is the wave frequency, $c$ is the
speed of light, and $\epsilon_{ij}$ are the elements of the dielectric
tensor, consisting of the susceptibilities of the different plasma
species: {$\mbox{\boldmath${\epsilon}$} = \textbf{1} +
  \mbox{\boldmath${\chi}$} ^i + \mbox{\boldmath${\chi}$} ^e$}. Here,
the indices $i$ and $e$ denote the ion and thermal electron
populations, respectively. We neglect the contribution of the runaway
electron population to the real part of the frequency.  In order to
make the calculation of the instability growth rate more convenient, we
rewrite the dispersion relation by introducing the cold plasma
formulas for the ion and thermal electron susceptibility tensor
elements in the high-frequency case of $\omega \gg \sqrt{m_e/m_i} \,
\omega_{ce}$. Equation~\eqref{eq:fulldisp} then becomes
\begin{align}
\label{eq:disp-eq}
&\omega^8 + \omega^6 C_1(k, \theta) + \omega^4 C_2(k, \theta) + \omega^2 C_3(k, \theta) + C_4(k, \theta) = 0,
\end{align}
where
\begin{align*}
C_1(k, \theta) &= -\left(2 k^2 c^2 + \omega_{ce}^2 + 3 \omega_{pe}^2\right),\\
\nonumber
C_2(k, \theta) &= k^4 c^4 + 2 k^2 c^2 (\omega_{ce}^2 + 2 \omega_{pe}^2) + \omega_{pe}^2 (\omega_{ce}^2 + 3 \omega_{pe}^2),\\
\nonumber
C_3(k, \theta) &= -\left[ k^4 c^4 (\omega_{ce}^2 + \omega_{pe}^2) + k^2 c^2 \omega_{pe}^2 (3/2 \omega_{ce}^2 + 2 \omega_{pe}^2 + 1/2 \omega_{ce}^2 \cos 2 \theta) + \omega_{pe}^6 \right],\\
\nonumber
C_4(k, \theta) &= 1/2 k^4 c^4 \omega_{ce}^2 \omega_{pe}^2 (1 + \cos 2 \theta),
\end{align*}
$\omega_{pe}$ is the electron plasma frequency and $\omega_{ce}$ is the
electron cyclotron frequency.  Equation \eqref{eq:disp-eq} is a
fourth order equation for $\omega^2$ giving four different branches of
electromagnetic waves, as described in Ref.~\cite{komarvarenna}. It
has been shown in Ref.~\cite{komarpop2012} that the two highest
frequency branches cannot be destabilized by the runaway
population. The remaining two branches, namely the electron-whistler
and the EXEL wave, can be destabilized but the EXEL wave was shown to
have a growth rate an order of magnitude higher than the
electron-whistler wave for a runaway distribution function relevant for near-critical electric field \cite{komarvarenna}.

\begin{figure}[htbp]
  \begin{center}
	\includegraphics[width=0.45\textwidth]{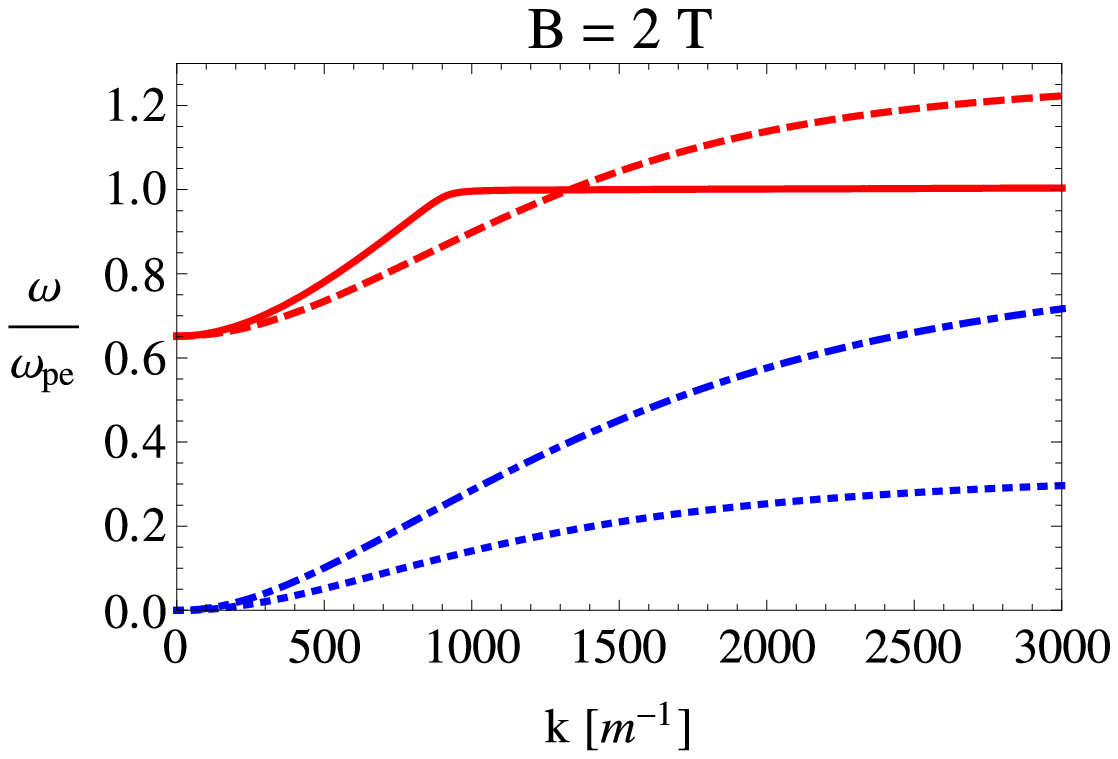}
	\includegraphics[width=0.45\textwidth]{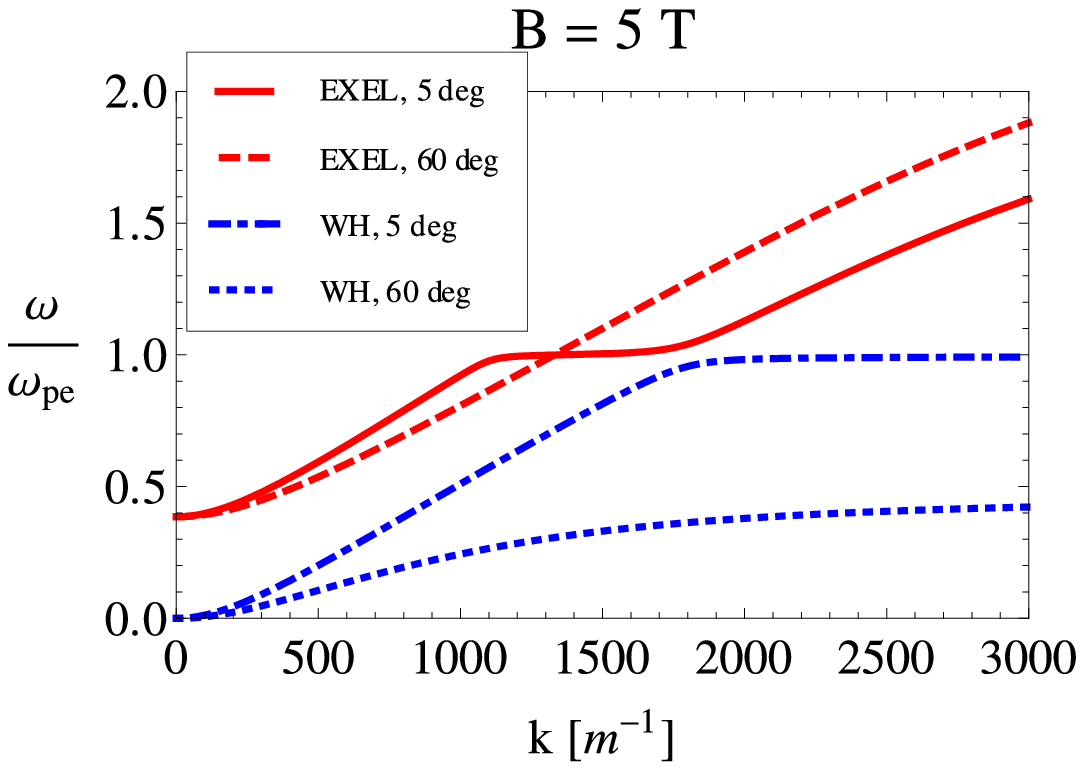}
	 \caption{Dispersion of the extraordinary-electron (EXEL) and the electron-whistler (WH) waves at different propagation angles for magnetic fields (a) $B=2\;\rm T$ and (b) $B=5\;\rm T$. The thermal electron density is $n_e = 5\cdot 10^{19}\;\rm m^{-3}$.}
\label{fig:disp_EXEL}
\end{center}\end{figure}

Figure \ref{fig:disp_EXEL} shows the dispersion of the EXEL and the electron-whistler waves for two propagation angles at two magnetic field values. For close-to-parallel propagation, both waves have wave number regions with approximately constant dispersion at the electron plasma frequency, while this feature gradually disappears for more oblique propagation directions.

\subsection{Linear growth rate}

By taking into account the contribution of runaway electrons to the
imaginary part of the frequency in the dispersion \eqref{eq:disp-eq},
the linear growth rate $\gamma_l$ of the EXEL wave is given by
Ref.~\cite{komarvarenna} as:
\begin{eqnarray}
\label{eq:gammal}
\frac{\gamma_l}{\omega^6 \left(\omega^2-\omega_{ce}^2\right)} = \Im
\left\{-\chi_{11}^r \left(\epsilon_{11} - {k^2 c^2}/{\omega^2} \right)
  \left(\epsilon_{33} - {k_\perp^2 c^2}/{\omega^2} \right) - 2
  \chi_{12}^r \epsilon_{12} \left(\epsilon_{33} - {k_\perp^2
      c^2}/{\omega^2} \right) -\right.\nonumber\\
  \chi_{22}^r \left[ \left(\epsilon_{11} - {k_\parallel^2 c^2}/{\omega^2} \right) \left(\epsilon_{33} - {k_\perp^2 c^2}/{\omega^2} \right) - {k_\parallel^2 k_\perp^2 c^4}/{\omega^4} \right] -\nonumber\\
\left.  \chi_{33}^r \left[\epsilon_{11}^2 - \epsilon_{11} \left({k^2
        c^2}/{\omega^2} + {k_\parallel^2 c^2}/{\omega^2} \right) +
    {k^2 k_\parallel^2 c^4}/{\omega^4} + \epsilon_{12}^2\right]
\right\} / F(\omega, k, \theta),
\end{eqnarray}
where
\begin{eqnarray}
F(\omega,k, \theta) = 8 \omega^7 + 6 \omega^5 C_1(k, \theta)  + 4 \omega^3 C_2(k, \theta) + 2 \omega C_3(k, \theta)
\end{eqnarray}
is the derivative of Eq.~\eqref{eq:disp-eq} with respect to $\omega$. 

The most important resonant interaction between the runaways and the
EXEL wave occurs when the wave frequency $\omega$ and wave-number $k$
are such that $\omega-k_\parallel v_\parallel = -\omega_{ce}/\gamma$,
where $v$, $\gamma$ and $\omega_{ce}$ are the velocity, relativistic
factor and the cyclotron frequency of the electrons taking part in the
interaction, respectively. This resonance is called the anomalous Doppler resonance.

In the case of the EXEL wave, the anomalous Doppler resonance occurs
with ultra-relativistic runaway electrons ($p \gg 1$, where $p=\gamma
v/c$ is the normalized momentum). In this region of the momentum space
the distribution function of the runaway electrons is highly
anisotropic. Meanwhile, the Cherenkov resonance $\omega-k_\parallel
v_\parallel = 0$ occurs with slightly relativistic runaways having
significantly lower normalized momentum ($p \approx 1$), for the same
wave frequency and wave number vector. For other resonances, such as
the Doppler resonance, the resonant momentum would be in the
negative region ($p < 0$). Thus, for a velocity distribution which is
sufficiently isotropic for low momentum, so that the Cherenkov
resonance can be neglected, and anisotropic for higher momentum, the
anomalous Doppler resonance will be dominant.

In the present analysis the effect of the Cherenkov resonance was neglected and a model for the ultra-relativistic runaway tail was used as initial electron distribution for the quasi-linear analysis. The distribution is given by
\begin{equation}
f_0(p_\|,p_\perp,t) = \frac{n_{r0} \alpha}{ 2 \pi c_Zp_\|} \exp\left( \frac{(E-1)t/\tau_c -
  p_\|}{c_Z} - \frac{\alpha p_\perp^2}{2 p_\|} \right) \left( \exp \left( \frac{p_\parallel - p_\mathrm{max}}{\sigma_p} \right) + 1 \right)^{-1},
\label{eq:initial}
\end{equation}
where the first part is the analytic secondary generation distribution
derived in Ref.~\cite{fuloppop2006}, valid for $E\gg 1$. In the above
equation, $E=e|E_\parallel|\tau_c/m_{e0} c$ is the normalized parallel
electric field (assumed to be constant in time), $m_{e0}$ is the
electron rest mass, $\tau_c=4\pi\epsilon_0^2m_{e0}^2 c^3/n_e e^4
\ln{\Lambda}$ is the collision time for relativistic electrons, $n_e$
is the background electron density,
$c_Z=\sqrt{3(Z+5)/\pi}\ln{\Lambda}$, $Z$ is the effective ion charge,
$\alpha=(E-1)/(Z+1)$ and $n_{r0}$ is the seed produced by primary
generation. In Eq.~\eqref{eq:initial} this form is supplemented by a
Fermi function imposing a gradual cut-off at high momentum around
$p_\mathrm{max}$ with a width of $\sigma_p$. This latter factor is
necessary to account for the maximum energy the electrons typically
reach, which is determined by the finite time duration of the
accelerating electric field \cite{Papp_PPCF_53_095004_2011} and the
energy loss due to close collisions \cite{Boozer}. In the present
paper, $p_\mathrm{max}=30$ (corresponding to an energy of $15\;\rm
MeV$) was chosen. This is the order of magnitude of the maximum
runaway electron energies typically observed in experiments, see
e.g. Figure 13 of Ref.~\cite{hollmann2013}. The width was chosen to be
$\sigma_p=1$. The runaway electron distribution in
Eq.~\eqref{eq:initial} is only valid for highly relativistic runaways,
and as such can only be used to calculate the resonant interaction
through the anomalous Doppler resonance.

\subsection{Most unstable wave and stability thresholds}

\begin{figure}[htbp]
  \begin{center}
	\includegraphics[width=0.45\textwidth]{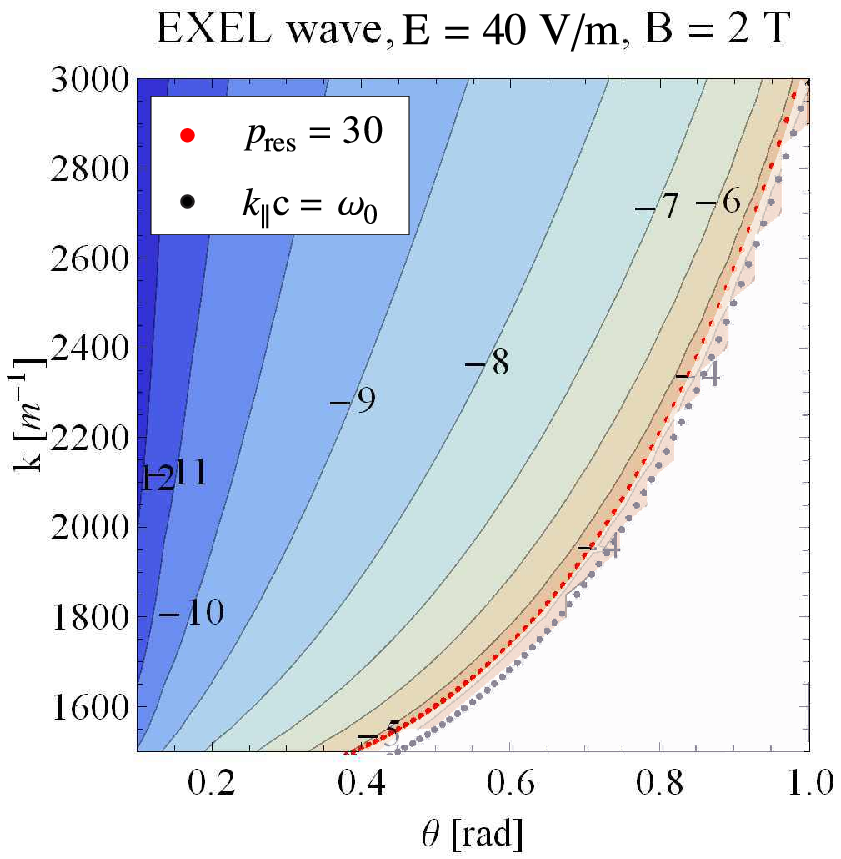}
	\includegraphics[width=0.45\textwidth]{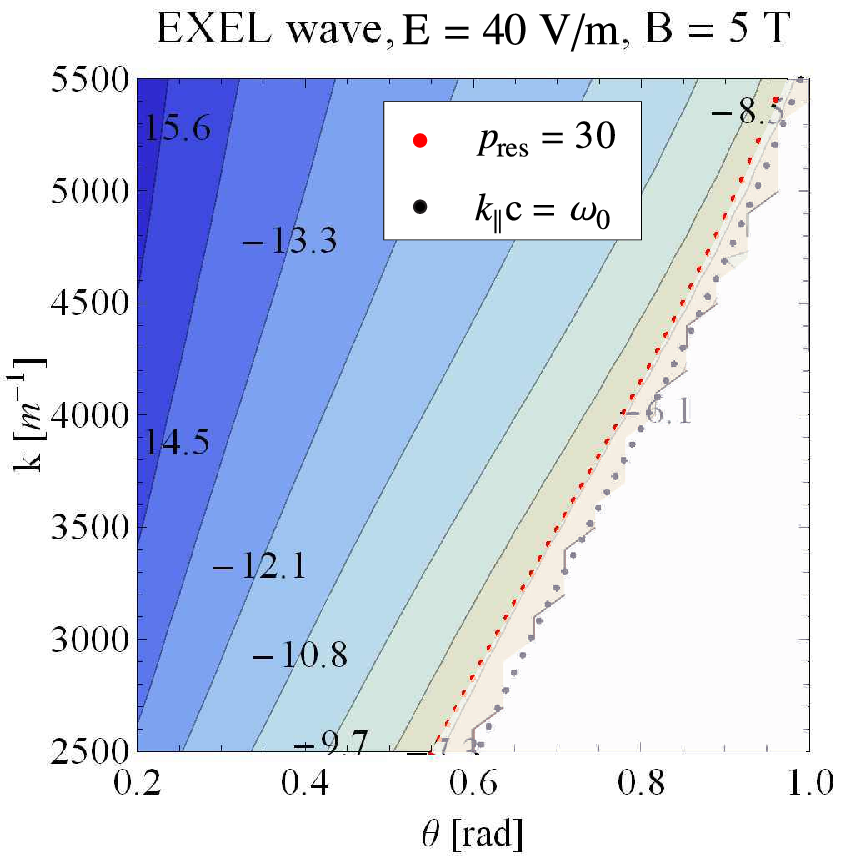}
	 \caption{Growth rate (contour lines) of the extraordinary electron wave ($\ln [\gamma_l / \omega_{ce}]$ is plotted) in the $k_\parallel c > \omega$ region for different magnetic fields (a) $B=2\;\rm T$ and (b) $B=5\;\rm T$. The grey dotted line shows where $k_\parallel c = \omega _0$, and the red dotted line where $p_{\textrm{res}}=p_{\textrm{max}}=30$. The parameters are electric field $E_\|=40\;\rm V/m$, thermal electron density $n_e = 5 \cdot 10^{19}$ $\text{m}^{-3}$ , runaway density $n_r = 3 \cdot 10^{17}$ $\text{m}^{-3}$ and effective ion charge $Z=1$.}
\label{fig:growth_EXEL}
\end{center}\end{figure}

\begin{figure}[htbp]
  \begin{center}
	\includegraphics[width=0.45\textwidth]{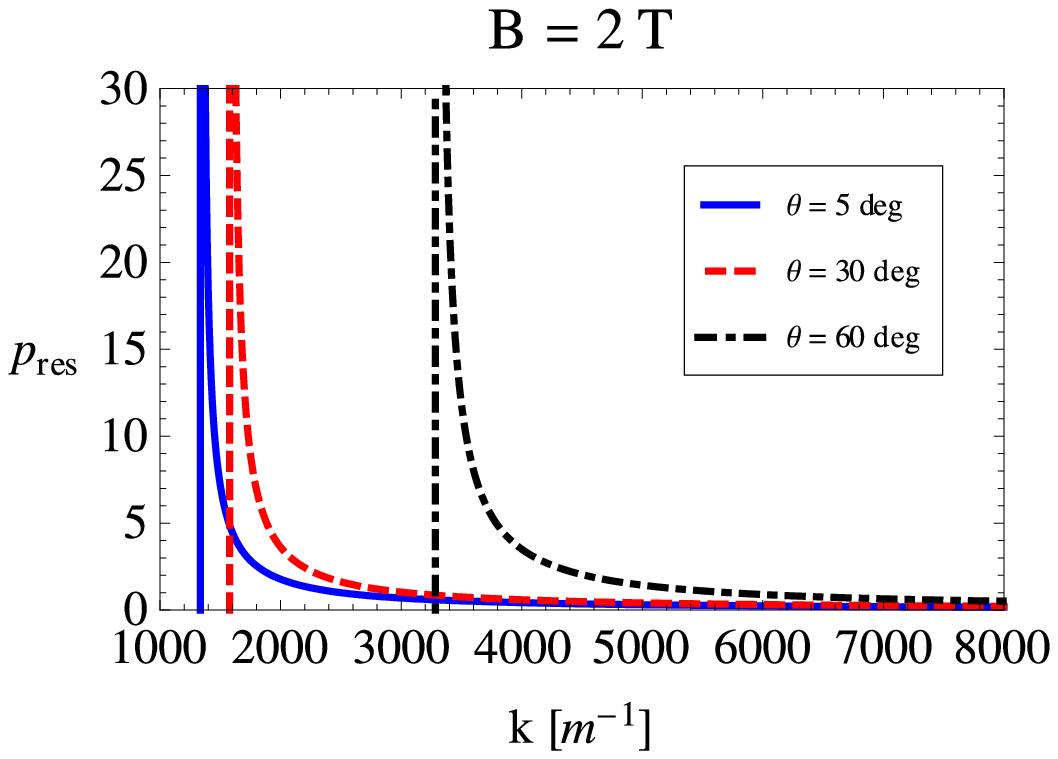}
	\includegraphics[width=0.45\textwidth]{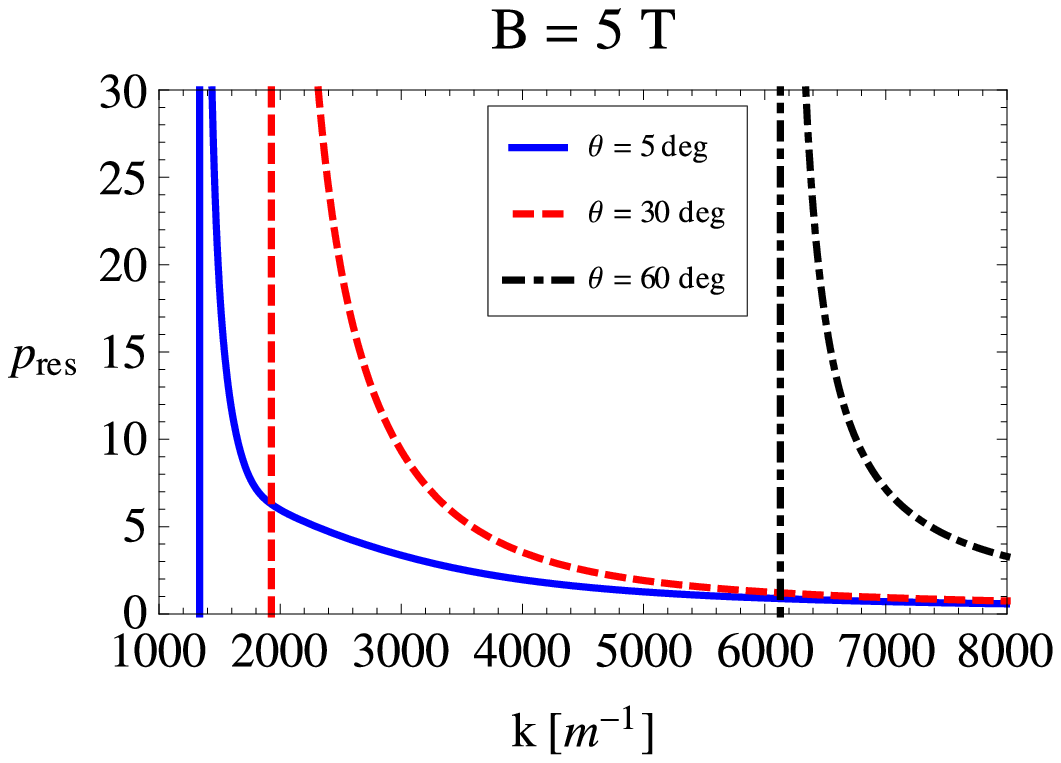}
        \caption{Resonant momentum for the EXEL wave as a function of
          the wave-number at different propagation angles ($\theta =
          5^{\circ}$, $30^{\circ}$ and $60^{\circ}$) for (a) $B = 2
          \;\rm T$ and (b) $B = 5 \;\rm T$ magnetic fields for the
          anomalous Doppler resonance. The thermal electron density is
          $n_e = 5\cdot 10^{19}\;\rm m^{-3}$.}
\label{fig:resonant_momentum}
\end{center}\end{figure}

The linear growth rate of the EXEL wave is calculated by substituting
the EXEL dispersion given by the second lowest frequency solution of
Eq.~\eqref{eq:disp-eq} and the runaway electron susceptibility
\cite{fuloppop2006,stix} into Eq.~\eqref{eq:gammal}. It is positive in
the whole wave number space, but the growth rate is highest in the
high wave number region, where $k_\parallel c > \omega$, see
Fig.~\ref{fig:growth_EXEL}. The growth rate increases as the
parameters get closer to the $k_\parallel c = \omega$ line. (Note,
that values closer to $k_\parallel c = \omega$ than the $p_{res}=30$
line in Fig.~\ref{fig:growth_EXEL} (red points) would only be valid
for a distribution function without the cut-off at
$p_{\mathrm{max}}=30$.)  

By approaching the $k_\parallel c = \omega$ line in the wave number
plane, the resonant momentum $p_\mathrm{res}$ of the runaways needed
for the destabilization of the wave increases rapidly. This is
illustrated in Fig.~\ref{fig:resonant_momentum} by showing curves
calculated at different wave propagation angles  $\theta$. This was
done by substituting the EXEL dispersion into the anomalous Doppler
resonance condition $\omega-k_\parallel v_\parallel =
-\omega_{ce}/\gamma$. The origin of the divergence at the $k_\parallel
c = \omega$ line can be understood by inserting $v_\parallel\approx c$
in the resonance condition.

Since the growth rate is increasing as we approach the $k_\parallel c
= \omega$ line (as seen on Fig.~\ref{fig:growth_EXEL}) and the closer
we are to $k_\parallel c = \omega$, the higher is the resonant
momentum $p_\mathrm{res}$, it follows that the resonant momentum of
the {\em most unstable} EXEL wave is close to the chosen maximum momentum
$p_\mathrm{res}\simeq p_\mathrm{max}$. However, we note that the exact
value of the chosen $p_\mathrm{max}$ does not have any drastic effect
on either the growth rate or the parameters of the resonant wave. The
reason for that is that the resonant wave parameters are not changed
significantly as $p_\mathrm{max}$ changes (as seen on
Fig.~\ref{fig:resonant_momentum}, the wave number $k$ is almost
the same whether we have e.g. $p_\mathrm{res} =30$ or
$p_\mathrm{res}=20$). Also no divergence in the growth rate is
observed when approaching the $k_\parallel c = \omega$. Therefore, the
order of magnitude of the growth rate for the most unstable wave is
the same, irrespective of the choice of $p_\mathrm{max}$.
However, due to the fact that the line corresponding to
$p_\mathrm{res} = p_\mathrm{max}$ nearly coincides with the contour
lines of the growth rate (as shown in Fig.~\ref{fig:growth_EXEL}), it
is not trivial to find the exact parameters of the most unstable
wave. Fortunately, for precisely the same reason, the value of the growth rate
or the number of runaways needed for the interaction are not affected significantly by the exact value of these parameters.

\begin{figure}[htbp]
  \begin{center}
	\includegraphics[width=0.45\textwidth]{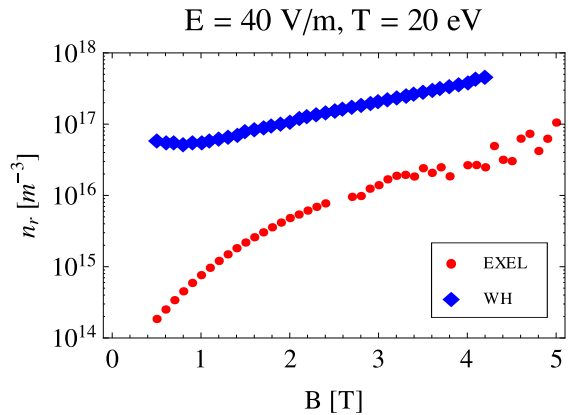}
	 \caption{Stability thresholds for the most unstable magnetosonic-whistler and EXEL waves in a strong electric field. The parameters are $T_e=20\;\rm eV$, $n_e = 5\cdot10^{19}$ $\text{m}^{-3}$, $Z=1$, $p_{\mathrm{max}}=30$, $L_r = 0.1\;\rm m$ (effective runaway beam radius \cite{pokolppcf2008}).}
\label{fig:stability}
\end{center}\end{figure}

Comparing the linear growth rate of the most unstable wave ($\gamma_l$)
to the collisional damping rate $\gamma_d=1.5 \tau_{ei}^{-1}$
\cite{brambilla} (where
$\tau_{ei}=3\pi^{3/2}m_{e0}^{2}v_{Te}^{3}\epsilon_{0}^2/n_iZ^2e^4\ln{\Lambda}$
is the electron-ion collision time), and the convective damping rate
$\gamma_v\equiv(\partial \omega/\partial k_\perp)/(4 L_r)$ (where
$L_r$ is the radius of the runaway beam \cite{pokolppcf2008}), gives
the linear stability threshold - the number of runaway electrons needed
for the destabilization of the
wave. %The stability threshold is orders
%of magnitude lower for the EXEL wave than for the electron-whistler
%wave in a low electric field (near-critical) case
%\cite{komarvarenna}. 
In the high electric field case studied in the present paper, the momenta
of the resonant runaways is expected to be higher than in
the corresponding near-critical case studied in
Ref.~\cite{komarvarenna}, and both
the most unstable EXEL and whistler waves therefore have lower frequencies. For the whistler wave this means that
instead of the high-frequency electron-whistler approximation, the magnetosonic-whistler wave \cite{fuloppop2006} - which also includes the ion susceptibilities in the dispersion relation - should be used.  The
stability thresholds for the EXEL and the magnetosonic-whistler wave
are shown in Fig.~\ref{fig:stability}. Note that the stability
threshold for the EXEL wave is significantly lower in this high
electric field case compared to the near-critical case studied in  Ref.~\cite{komarvarenna}. Here, the electric field was chosen to be $40 \;\rm
V/m$.

For the reference scenario of Fig~\ref{fig:stability} at $B = 2 \;\rm
T$, the parameters of the most unstable wave are: wave-number
$k_m\simeq 4900 \;\rm m^{-1}$, wave vector angle $\theta_m=1.2$ and
frequency \mbox{$\omega_0 \simeq 5.1 \cdot 10^{11}\;\rm s^{-1}$}. Although the
EXEL wave dispersion is generally quite complicated, in the vicinity of
the most unstable wave the dispersion can be approximated by a linear trend in wave
number $k$ and a linear dependence in $\sin{\theta}$:
\begin{equation}
\omega_\mathrm{fit}(k,\theta) = C_\omega \omega_{pe}+C_k kc+C_\theta \omega_{ce}\sin{\theta},
\label{eq:omega-fit}
\end{equation}
where $C_\omega = 0.92$, $C_k = 0.011$ and $C_\theta = 0.35$ around
the most unstable wave in the reference scenario. The values of $C_\omega$ and $C_\theta$
tend to be quite robust with respect to changes in
plasma density and magnetic field strength; a variation of only about 5\% is  observed
for a change in the plasma parameters of roughly 20\%. $C_k$ increases
very strongly with increasing magnetic field, but remains
almost insensitive to changes in the background electron density. This
parameter gives a relatively small contribution to
$\omega_\mathrm{fit}$, so the fit is considered to be quite good in
the close vicinity of the most unstable wave in the reference scenario.
The fit also reproduces some of the dominant changes in the dispersion due to changes in the 
plasma parameters. However, in the region of interest (which
is quite large due to the large spectral range of the waves
destabilized in the quasi-linear interaction), $\omega_\mathrm{fit}$ deviates significantly from the exact dispersion. In the remainder of this paper, the exact dispersion given by the solution of Eq.~\eqref{eq:disp-eq}, will be used.

\section{Quasi-linear development of the extraordinary-electron wave instability}
\label{sec:III}

In the framework of quasi-linear theory, the evolution of the
distribution function of the electrons is given by a diffusion
equation in phase space, and the rate of growth of wave-energy is
equal to the difference between the linear growth rate and the damping rates,
$\gamma_{k}=\gamma_{l}-\gamma_{d}-\gamma_{v}$. The analysis of the
dynamics of the interaction of runaway electrons with the EXEL wave can
be performed similarly to that of the magnetosonic-whistler wave in
Ref.~\cite{pokolppcf2008}. Only the dispersion relation and the
polarization of the wave are different in this case, but as we will
show, this proves to have a significant effect on the temporal
evolution of the instability. The evolution of the runaway
distribution in the general case is given by \cite{stix}:
\begin{equation}
\frac{\partial f}{\partial t}=\frac{\pi e^2}{m_{e0}^2 c^2} \sum_{n=-\infty}^{\infty}\int d^3 k\ \hat{\Pi}\, p_\perp \delta(\omega_k-n \Omega -k_\parallel p_\parallel c/\gamma)\,\frac{|\psi_{n,k}|^2}{\omega^2} p_\perp \hat{\Pi} f ,
\end{equation}
where $\Omega=eB/m_e=\omega_{ce}/\gamma$,
\begin{equation}
\hat{\Pi}=\frac{\omega-k_\parallel p_\parallel c/\gamma}{ p_\perp}\frac{\partial}{\partial p_\perp}+\frac{k_\parallel c}{\gamma}\frac{\partial }{\partial p_\parallel},
\label{pi}
\end{equation}
\begin{equation}
|\psi_{n,k}|^2= \left|E_{kx} \frac{n}{z}J_n+ i E_{ky}J_n^\prime+\frac{p_\parallel}{p_\perp}E_{kz} J_n\right|^2,
\label{psink}
\end{equation}
$E_{kx}$, $E_{ky}$, $E_{kz}$ are the components of the spatial Fourier transform of the electric field and $J_n (z)$ is the Bessel function of the first kind and of order $n$, with the argument $z=k_\perp p_\perp c/\omega_{ce}$.  Using the polarization for the EXEL wave
\begin{equation}
\left(e_x , e_y , e_z \right) = \left( 1 \,,\; -i \frac{\omega_{pe}^2 \omega_{ce}/\omega}{\omega^2 - k^2 c^2 - \omega_{ce}^2 - \omega_{pe}^2 + k^2 c^2 \omega_{ce}^2/\omega^2} \,,\; \frac{k_\parallel k_\perp c^2}{\omega_{pe}^2 + k_\perp^2 c^2 - \omega^2} \right),
\end{equation}
where $\omega=\omega_{\mathrm{EXEL}}(k,\theta)$, equation (\ref{psink}) gives
\begin{equation}
|\psi_{n,k}|^2= |E_{kx}|^2\left| \frac{nJ_n}{z}+ \frac{J_n^\prime\omega_{pe}^2 \omega_{ce}/\omega}{\omega^2 - k^2 c^2 - \omega_{ce}^2 - \omega_{pe}^2 + k^2 c^2 \omega_{ce}^2/\omega^2}+\frac{p_\parallel}{p_\perp}\frac{k_\parallel k_\perp c^2J_n}{\omega_{pe}^2 + k_\perp^2 c^2 - \omega^2} \right|^2 .
\label{psi}
\end{equation}
%The last term in $|\psi_{n,k}|^2$ can be neglected since for the
%considered instability, with wave numbers $k_\perp\sim \omega_{pi}/2
%v_A$, $k_\parallel \sim 2 \omega_{ce}/c c_z$, the condition $1/z\gg
%p_\parallel k_\parallel k_\perp c^2/p_\perp \omega_{pe}^2$ leads to
%$c_z/z\gg p_\parallel/p_\perp$, which is satisfied in the parameter
%regime of interest, as it can be shown {\it a posteriori}.

The wave instability is driven by the anisotropy of the runaway distribution via the anomalous Doppler resonance $n=-1$. For $z\ll 1$ the Bessel function can be expanded as $J_{-1}=-J_1\simeq -z/2$ in $|\psi_{-1,k}|^2,$ and using $|E_k|^2=|E_{kx}|^2(|e_x|^2+|e_y|^2+|e_z|^2)$ we obtain
\begin{equation}
|\psi_{-1,k}|^2=\frac{|E_{k}|^2}{4}\frac{\left|1-\frac{\omega_{pe}^2 \omega_{ce}/\omega}{\omega^2 - k^2 c^2 - \omega_{ce}^2 - \omega_{pe}^2 + k^2 c^2 \omega_{ce}^2/\omega^2}-p_\parallel\frac{k_\parallel k_\perp^2 c^3}{\omega_{ce}(\omega_{pe}^2 + k_\perp^2 c^2 - \omega^2)}\right|^2}{|e_x|^2+|e_y|^2+|e_z|^2}  \equiv \frac{|E_{k}|^2}{4} P(\omega, k, \theta, p_\parallel).
\end{equation}
The quasi-linear equation for the runaway distribution becomes
\begin{equation}
\frac{\partial f}{\partial t}=\frac{\pi e^2}{m_{e0}^2 c^2}\int d^3k\ \hat{\Pi}\,\frac{|E_{k}|^2}{4} P(\omega, k, \theta, p_\parallel)\, \frac{p_\perp^2 }{\omega^2 }\,\delta(\omega+\Omega-k_\parallel p_\parallel c/\gamma)\,\hat{\Pi}\,f,
\label{deq}
\end{equation}
and if we assume $k_\parallel v_\perp \partial f/\partial p_\parallel
\ll \Omega \partial
f/\partial p_\perp$,  Eq.~(\ref{deq}) simplifies to
a diffusion equation
\begin{equation}
\frac{\partial f(p_\perp, p_\parallel, t)}{\partial
  t}=\frac{1}{\gamma p_\perp}\frac{\partial}{\partial p_\perp}\left(
  \frac{p_\perp D(p_\perp,p_\parallel,t)}{\gamma}\frac{\partial f(p_\perp, p_\parallel, t)}{\partial p_\perp}\right)
\label{eq:diffeq}
\end{equation}
with 
\begin{equation}
\label{eq:diffop}
D(p_\perp,p_\parallel,t) = \frac{\pi e^2 \omega_{ce}^2}{2 \epsilon_0 m_{e0}^2 c^2} \int d^3 k \frac{W_k(t)}{\omega^{2}} P(\omega, k, \theta, p_\parallel) \, \delta(\omega+\Omega-k_\parallel p_\parallel c/\gamma),
\end{equation}
where $W_k(t)=\displaystyle\frac{\epsilon_0}{2} |E_k(t)|^2$ is the
  spectral energy of the wave. The assumption $k_\parallel v_\perp
  \partial f/\partial p_\parallel \ll \Omega \partial f/\partial
  p_\perp$ is valid when $z k_\parallel /k_\perp \ll
  (\partial f / \partial p_\perp)/(\partial f / \partial p_\parallel)$, and is satisfied due to the ordering
  $z\ll 1$, $k_\parallel \sim k_\perp$ and $(\partial f / \partial p_\perp)/(\partial f / \partial p_\parallel) \gg 1$.

The time variation of the spectral energy of the wave is determined by the differential equation \cite{pokolppcf2008}:
\begin{equation}
\label{eq:wk} 
\frac{dW_k}{dt}=2 \gamma_k(t) W_k,
\end{equation}
with the initial condition $W_{k0}=W_k(t=0)= e T_e/2$, which is the thermal fluctuation level.

\subsection{Numerical solution}
\label{sec:diffusion}

Assuming a beam-like velocity distribution $\gamma\simeq p_\parallel$ and  introducing all terms containing $p_\parallel$ in (\ref{eq:diffeq}) into the diffusion operator
\begin{equation}
\label{eq:diffop2}
\tilde{D}(p_\parallel,t)=\frac{\pi e^2 \omega_{ce}^2}{2 \epsilon_0 m_{e0}^2 c^2}\frac{1}{p_\parallel^2}\int d^3 k\frac{W_k(t)}{\omega^{2}} P(\omega, k, \theta, p_\parallel) \delta(\omega+\omega_{ce}/p_\parallel-k_\parallel c),
\end{equation}
we obtain a diffusion equation for $f$ in which $\tilde{D}(p_\parallel,t)$ is independent of $p_\perp$. Introducing a dimensionless time
\begin{equation}
\tau{(p_\parallel,t)}= \int_0^t dt^\prime \tilde{D}(p_\parallel,t^\prime),
\label{eq:tau}
\end{equation}
the diffusion equation (\ref{eq:diffeq}) takes the form:
\begin{equation}
\frac{\partial f}{\partial \tau}=\frac{1}{p_\perp}\frac{\partial}{\partial p_\perp} p_\perp \frac{\partial f}{\partial p_\perp}\ ,
\label{eq:diffeq3}
\end{equation}
and with the initial condition (\ref{eq:initial}) the solution according to \cite{pokolppcf2008} is
\begin{equation}
f(p_\perp,p_\parallel, t)=\frac{n_{r0} \alpha}{ 2 \pi c_Z\phi(p_\parallel, t)} \exp\left( \frac{(E-1)t/\tau_c -  p_\|}{c_Z} - \frac{\alpha p_\perp^2}{2\phi(p_\parallel, t)} \right) \left( \exp \left( \frac{p_\parallel - p_\mathrm{max}}{\sigma_p} \right) + 1 \right)^{-1},
\label{eq:sol}
\end{equation}
where $\phi(p_\parallel, t)= 2\alpha \tau(p_\parallel,t)+p_\parallel$.

This formula gives the evolution of the runaway distribution as a
function of the dimensionless time, $\tau (p_\parallel,t)$. 
This enables us to create a numerical code which only has to
solve for $\tau$ in each time step for every $p_\parallel$ value in a
certain region. In order to calculate $\tau$, we need to evaluate the
integral in Eq.~(\ref{eq:diffop2}). Due to the azimuthal symmetry of
the system, Eq.~\eqref{eq:diffop2} can be written on the form
\begin{equation}
\tilde{D}(p_\parallel,t) \sim 2 \pi \int dk \, d\theta \, k^2 \sin\theta \; G(k,\theta) \; \delta \left(\omega(k,\theta)+\frac{\omega_{ce}}{p_\parallel}-k c \cos \theta \right),
\end{equation}
where $G(k,\theta)$ is a function incorporating all dependences on the wave number and the propagation angle (other than the delta function and the Jacobian). We can evaluate the integral in $k$ and arrive at
\begin{equation}\label{eq:diffusion_coeff}
\tilde{D}(p_\parallel,t) = \frac{\pi^2 e^2 \omega_{ce}^2}{\epsilon_0 m_{e0}^2 c^2} \int d \theta \left[ \frac{W_k(t)}{p_\parallel^2 \omega^2 (k,\theta)} P(\omega, k, \theta, p_\parallel) \frac{k^2 \sin\theta}{\left| \frac{d\omega}{dk} - c \cos \theta \right|} \right]_{k=k_{\mathrm{res}}}\!\!\!\!\!\!\!\!\!\!\!\!\!,
\end{equation}
where $k_{\mathrm{res}}$ is the solution of the resonance condition
$\omega(k,\theta)+\omega_{ce}/p_\parallel-k c \cos \theta = 0$, which can be calculated numerically. From
now on we will refer to  $k_{\mathrm{res}}$ as the \emph{resonant curve}.

The numerical solution of the quasi-linear equations proceeds as
follows. For each time step and for each parallel momentum
the linear growth rate of the EXEL wave is calculated along the
resonant curve. Based on this, the wave energy is determined (with the
initial condition being the level of thermal fluctuations). Then the
diffusion coefficient is calculated by integrating along the resonant
curve, as prescribed by Eq.~(\ref{eq:diffusion_coeff}). Finally, the
diffusion coefficient is integrated in time to yield the dimensionless
time $\tau$, which in turn gives the distribution function for the
runaways.

\begin{figure}[htbp]
\begin{center}
\includegraphics[width=1\linewidth]{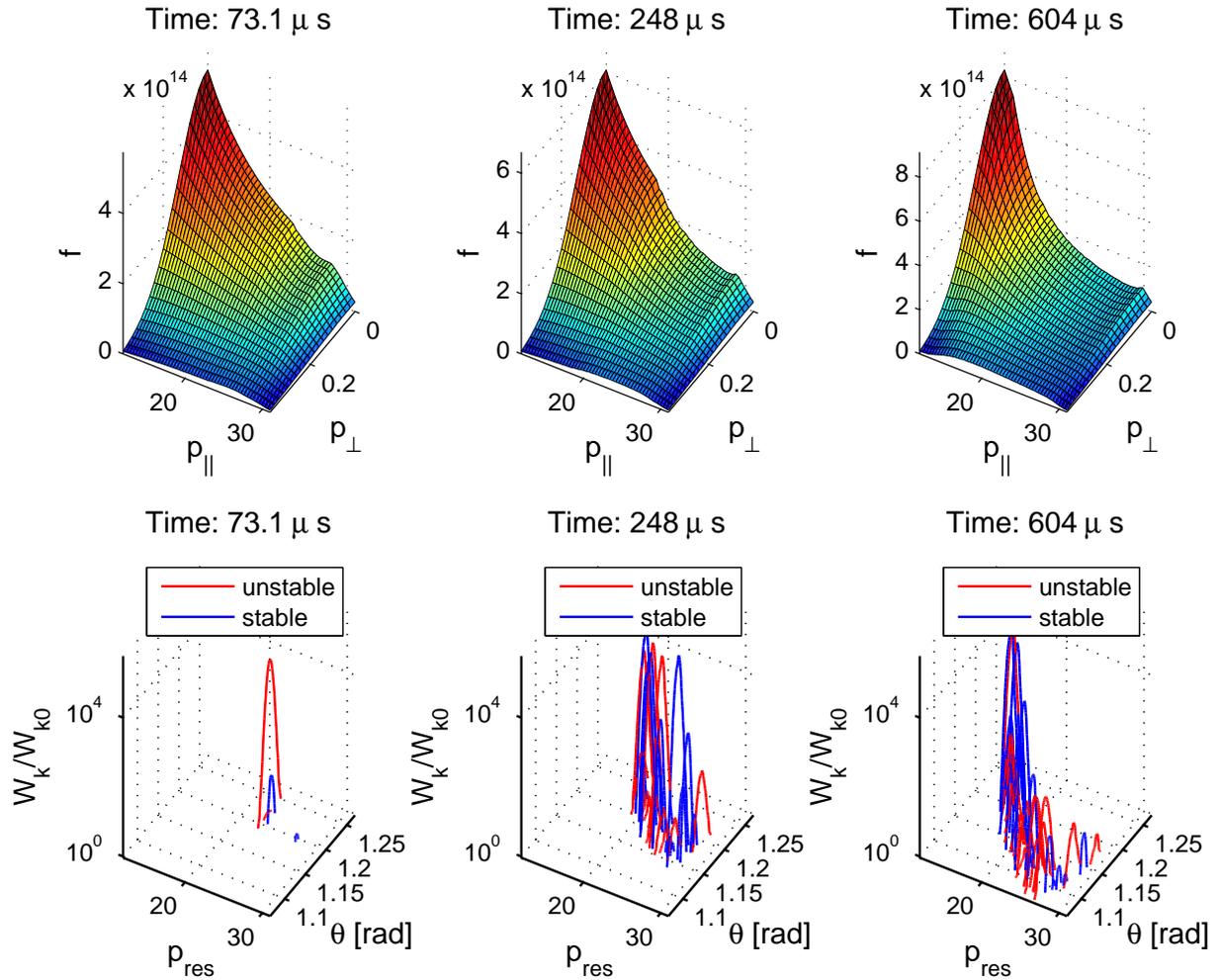}
\end{center} \vspace{-10mm}
\caption{\small \it \noindent Quasi-linear evolution of the runaway
  distribution and the wave energy of the EXEL wave at consecutive times. Red and blue
  lines correspond to the wave energy along the $k_\mathrm{res}$
  resonant curve as a function of $\theta$ for a certain
  $p_\mathrm{res}$ resonant momentum. The displayed time corresponds to the time elapsed
  since the first destabilization of the most unstable wave. The parameters
  correspond to the JET-like reference scenario: $B=2\;\rm T$, $n_e = 5 \cdot 10^{19}$ $\text{m}^{-3}$, $Z=1$,
  $T_e=20\;\rm eV$.}
	\label{fig:changing_JET}
\end{figure}

A reference scenario was chosen with the following JET-like parameters: magnetic field $B = 2 \;\rm
T$, thermal electron density $n_e = 5 \cdot 10^{19} \;\rm m^{-3}$, post-disruption background temperature $T = 20 \;\rm eV$ and electric field $E = 40 \;\rm V/m$. The quasi-linear
effect in this case is shown in Fig.~\ref{fig:changing_JET}, and can
be characterized as the following cycle. After some initial time the
runaway density reaches the critical value and the EXEL wave is
destabilized. During the evolution, the energy of the wave grows to
a certain point where the runaway distribution is affected by the wave
and the resonant electrons around $p_\parallel \sim 25$ are
pitch-angle scattered. This causes the wave energy to decrease while
the distribution is unaffected for some time. As the number of
runaways continues to grow on a longer time scale, the number of
runaways reaches the critical value again and the wave is destabilized
for a second phase of isotropization.

The effect seen in Fig.~\ref{fig:changing_JET} is the result of
several such wave destabilization cycles. During these cycles, the
part of the runaway distribution that is affected by the interaction  
(the resonant region) is spread out and the effects due to the individual cycles
accumulate to cause a significant pitch-angle scattering of the
runaways. This extension of the affected region can also be observed
by looking at the parameters of the resonant waves. Although the
propagation angles do not change significantly, the wave number region
affected becomes broader due to the broader interaction region in momentum space.

\section{Parametric dependencies}
\label{sec:IV}
        
It is instructive to examine the quasi-linear effect for a wider range of plasma parameters. For different magnetic fields, electric fields, background temperatures and thermal electron densities, we have looked at the differences in the final runaway distribution about $90 \;\rm \mu s$ after the destabilization of the EXEL wave by varying one parameter at a time. This time duration is not enough for the EXEL wave to cause such a large effect on the distribution function as shown in Fig.~\ref{fig:changing_JET}, however the first stage of the isotropization is clearly visible, allowing a characterization of the influence of the various parameters on the dynamics of the interaction.

\begin{figure}[htbp]
\begin{center}
\includegraphics[width=1\linewidth]{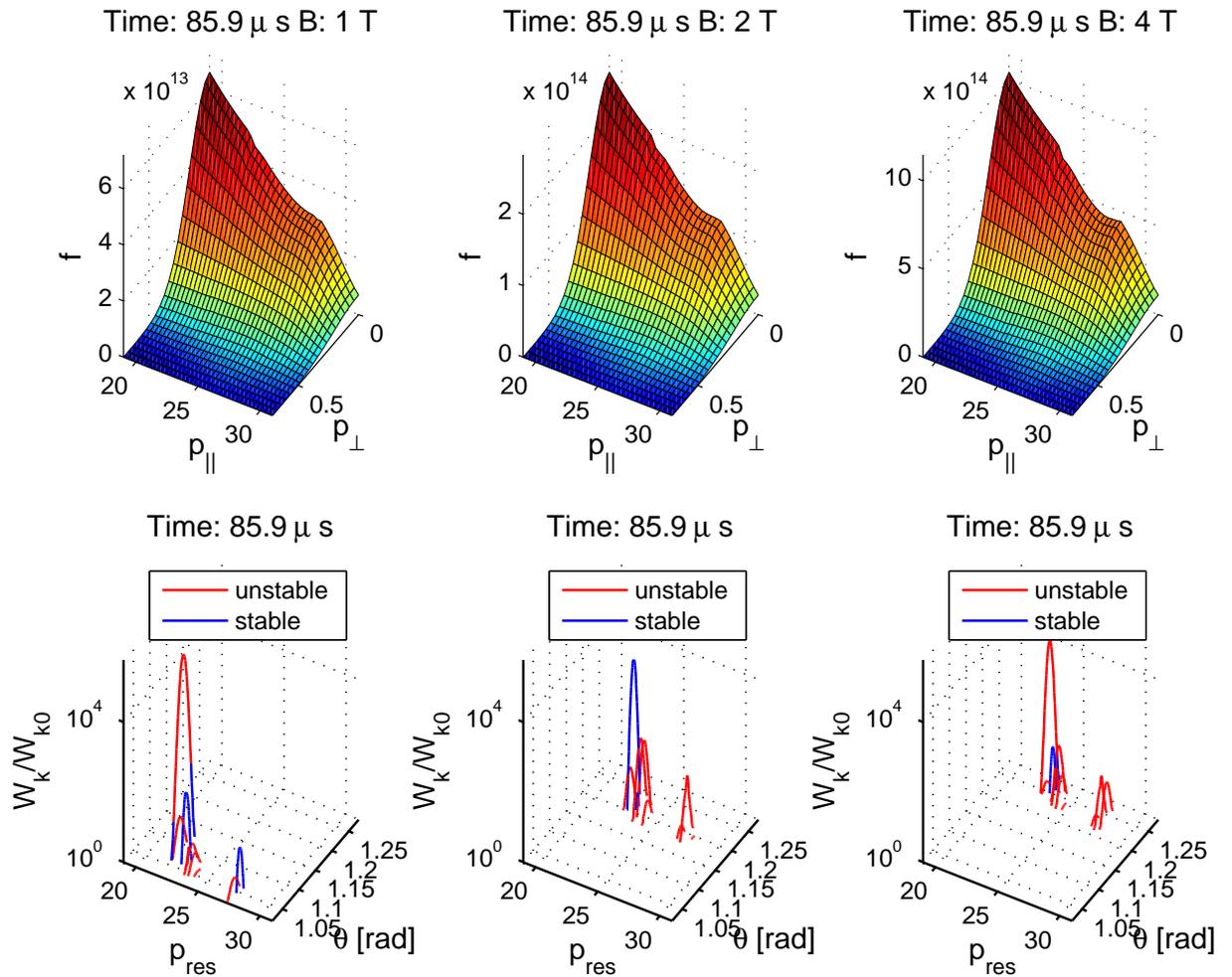}
\end{center} \vspace{-10mm}
\caption{\small \it \noindent Quasi-linear evolution of the runaway distribution and the wave energy at magnetic fields (a) $B=1\;\rm T$, (b) $B=2\;\rm T$ and (c) $B=4\;\rm T$, $85.9 \;\rm \mu s$ after the first wave destabilization. The parameters not displayed correspond to the reference values.}
	\label{fig:changing_B}
\end{figure}

An example, where the magnetic field strength was varied, is shown in
Fig.~\ref{fig:changing_B}. The figures for different magnetic fields are qualitatively similar; there are only two pronounced
differences. The first is that a significantly larger number of
runaway electrons is necessary for the destabilization of the EXEL wave
for high magnetic field strengths, as indicated by the vertical axes of the distribution function plots. This agrees with the trend shown
in Fig.~\ref{fig:stability}. The other significant difference is the
smaller angle of wave propagation at weak magnetic field, although the
difference of about 0.2 radians is not particularly large. Upon closer inspection, the extent of isotropization due to the EXEL wave seems slightly larger for stronger magnetic fields, but the difference is not significant.

A quantitative analysis has also been performed regarding the change in the quasi-linear evolution due to changes in the plasma parameters. The results are summarized in Table~\ref{table:bnet}, which shows the value of the following characterizing parameters: 
\begin{itemize}
\item $p_{\parallel}^{m}$,  the momentum resonant with the most unstable wave
\item $n_{r1}$, the runaway density at momentum $p_{\parallel}^{m}$ integrated over $p_{\perp}$ at the time of the first wave destabilization. $n_{r1}$ is thus the 'threshold linear density'
\item $W_{max}$, the maximum wave energy over the $85.9 \;\rm \mu s$ duration of the simulation
\item $\tau$, the parameter characterizing the extent of velocity space diffusion (calculated according to Eq.~\ref{eq:tau}) at $p_{\parallel}^{m}$ at the end of the simulation 
%after $85.9 \;\rm \mu s$ of interaction
\item $k_m$, the wave-number of the most unstable wave
\item $\theta_m$, the propagation angle of the most unstable wave
\end{itemize}

\begin{table}[htbp]
\begin{tabular}{|c|c|c|c|c|c|c|c|c|c|c|}
\hline
&\multicolumn{2}{|c|}{$B\;\rm [T]$ }&\multicolumn{2}{|c|}{$n\;\rm [m^{-3}]$ }&\multicolumn{2}{|c|}{$E\;\rm [V/m]$ }&\multicolumn{2}{|c|}{$T\;\rm [eV]$ }&\textbf{Reference scenario}\\ 
\cline{2-10}
&1&4&$2 \cdot 10^{19}$&$1 \cdot 10^{20}$&20&80&10&50&\textbf{$2\;\rm T$, $5\cdot 10^{19} \;\rm m^{-3}$}\\

&&&&&&&&&\textbf{ $20 \;\rm eV$, $40\;\rm V/m$}\\
\hline
$p_{\parallel}^{m}$&25.9&26.2&26.1&25.8&25.8&26.1&26.0&26.0&\textbf{26.0}\\
\hline
$n_{r1} \;[10^{13}\;\rm m^{-3}]$&0.8&13&2.4&4.4&4.6&2.3&5&2.3&\textbf{3.1}\\
\hline
$W_\mathrm{max} \; [10^{-12}\;\rm J]$&7.7&0.7&0.2&14&3.3&2.5&7&1.1&\textbf{2.7}\\
\hline
$\tau \;[10^{-3}]$&3.1&4.5&1.8&6.1&3.0&4.5&4.4&3.0&\textbf{3.6}\\
\hline
$k_{m} \; [10^{3}\;\rm m^{-1}]$&2.9&8.5&6.4&4.2&4.0&5.9&4.5&5.3&\textbf{4.9}\\
\hline
$\theta_{m}\;\rm [rad]$&1.06&1.24&1.34&1.04&1.13&1.27&1.17&1.23&\textbf{1.20}\\
\hline
\end{tabular}
\caption{\small \it \noindent Characteristic parameters of the quasi-linear interaction for different values of the magnetic field, thermal electron density, electric field and background temperature at a fixed time (85.9 $\mu$s) after the first wave destabilization. In each column, only the parameter indicated by the column heading was changed - the remaining parameters where those of the reference scenario.}
\label{table:bnet}
\end{table}

From Tab.~\ref{table:bnet} we can infer that the most unstable
momentum of the runaways is not affected significantly by the change
of the plasma parameters through the quasi-linear evolution - it is
close to the $p_\mathrm{max}$ cut-off value introduced in the initial
distribution function (Eq.~\ref{eq:initial}).

The changes due to variations in the magnetic field already described in the discussion of Fig.~\ref{fig:changing_B} are also visible in the table, with the
additional observation that larger magnetic fields shift the resonant
EXEL waves towards larger wave numbers. On the other hand the maximum
wave energy is an order of magnitude higher for lower magnetic field,
and is obtained by the most unstable wave during the first phase of
isotropization (whereas Fig.~\ref{fig:changing_B} shows a later time instant).

The dominant effect of a change in the density is a modification to the strength of
the collisional damping. Accordingly, a higher density means a higher critical runaway density ($n_{r1}$). On the other hand, the quasi-linear
diffusion is significantly faster for high densities and the wave energies are significantly higher.

Increasing the accelerating electric field results in a decrease of
the critical runaway density needed for the destabilization, which is
explained by the increasing anisotropy of the runaway beam. There is no substantial effect on the other parameters in Tab.~\ref{table:bnet}.

The background plasma temperature enters through the collisional
damping, so $n_{r1}$ increases with decreasing temperature. It is a general observation that a higher
threshold runaway density is accompanied by a higher wave energy. The only exception is modifications to the magnetic field strength, where the trend is the opposite.

In summary, the EXEL wave is expected to be destabilized in plasmas where the density and temperature are not too low, and where the magnetic field is weak. These conditions could be fulfilled in e.g. the thermal quench phase of tokamak disruptions, especially if an anisotropic fast electron
population (due to for instance lower hybrid or electron cyclotron resonance heating) is present just before the disruption. The parameters of the wave
remain in the $k_m \sim 3 - 8.5 \cdot 10^3 \;\rm m^{-1}$ and $\theta_m
\sim 1 - 1.3 \;\rm rad$ region, and the largest difference in wave-numbers
is caused by changes to the magnetic field strength. At the same time, the
spectral energy of the wave is on the order of $10^{-12}
- 10^{-11} \;\rm J$, making the direct detection of the wave
practically impossible.

\section{Impact on synchrotron radiation spectrum}
\label{sec:V}

Due to the low energy of the destabilized EXEL wave in our simulations
($W_\mathrm{max}\aplt 10^{-11}$ J, see Tab. \ref{table:bnet}), the
resonant interaction between the runaway electrons and the EXEL wave
are likely to be hard to detect directly. One possible way to infer
the presence of the interaction is to look at the spectrum of the
synchrotron radiation emitted by the highly relativistic runaways as a
consequence of their toroidal and gyro-motion. The emitted synchrotron
power is highly dependent on both the energy and pitch of the emitting
particle (it scales roughly as $P\propto \gamma^2 (v_{\perp}/v_{\|})^2
$ \cite{jaspers2001}), and for this reason pitch-angle scattering of
the runaways alters their synchrotron spectrum. The biggest effect of
the interaction with the EXEL wave is expected among the most
energetic runaways, but these are also the most strongly emitting
particles in terms of synchrotron radiation. Therefore, the
wave-particle interaction can result in a substantial change in the
synchrotron spectrum \cite{adamsyrup}.

The average synchrotron power emitted per runaway particle at a specific wavelength $\lambda$ can be calculated as a convolution of the distribution function with the synchrotron emission from a single particle: 
\begin{equation}
P(\lambda,t)=\frac{2\pi}{n_{r}(t)}\int_{S_{RE}}f(p_\|,p_\perp,t)\, \mathcal{P}(p_\|,p_\perp,\lambda)\, p_\perp\mbox{d}p_\|\,\mbox{d}p_\perp\ , \label{eq:total_emission}
\end{equation}
where $f$ is the momentum-space distribution of electrons, $\mathcal{P}$ describes the synchrotron emission and $S_{RE}$ is the runaway region in momentum space \cite{adamsyrup}. The synchrotron power radiated by a highly relativistic particle in a toroidal plasma was derived in Ref. \cite{pankratov} and is given by
\begin{align}
\mathcal{P}(\lambda)= c_P&\left\{ \int_{0}^{\infty}\!\!\!g(y)\, J_{0}\left(a \xi y^{3}\right)\sin\left(h(y)\right)\dy\right.\, -\,4a\!\int_{0}^{\infty}\!\!\! y\, J'_{0}\left(a \xi y^{3}\right)\cos\left(h(y)\right)\dy\, -\frac{\pi}{2} \Bigg\} ,\label{eq:fullPank}
\end{align}
where $c_P = ce^{2}/(\epso\lambda^{3}\gamma^{2})$, $a=\eta / (1+\eta^{2})$, $g(y)=y^{-1}\!+2y$, $h(y)=3\xi(y+y^{3}/3)/2$, 
% \mbox{$\xi = 4\pi R/(3\lambda\gamma^{3}\sqrt{1+\eta^{2}})$},
\begin{equation}
\xi =\frac{4\pi}{3}\,\frac{R}{\lambda\gamma^{3}\sqrt{1+\eta^{2}}}\ ,
\end{equation}
\begin{equation}
\eta=\frac{eBR}{\gamma m_e}\,\frac{\vp}{\vpa^{2}}\simeq\frac{\omega_{c}R}{\gamma c}\,\frac{\vp}{\vpa}\ , 
\end{equation}
$R$ is the tokamak major radius, $\gamma$ is the relativistic mass
factor, $J_\nu (x)$ is the Bessel function and $J_\nu'(x)$ its
derivative. Eq.~\eqref{eq:fullPank} takes the drifts stemming from the
curvature and gradient of the magnetic field into account and is valid
when $p_{\|} \gg p_{\perp}$, $c/\lambda \gg \omega_{ce}$, and when the
aspect ratio is large. It depends on the particle energy and pitch
through the parameters $\gamma$ and $\eta$, respectively. Due to their
structure, the integrands in Eq.~\eqref{eq:fullPank} are highly
oscillatory and evaluating the integrals can be numerically demanding.
Here we use a Matlab routine called SYRUP (SYnchrotron emission from
RUnaway Particles), also used to obtain the results in
Ref.~\cite{adamsyrup}, to calculate the synchrotron spectrum from the
normalized distribution before and after the onset of the resonant
interaction between the runaway distribution and the EXEL wave. SYRUP
implements Eqs.~\eqref{eq:total_emission} and \eqref{eq:fullPank}.

\begin{figure}[htbp]
\begin{center}
\includegraphics[width=0.5\linewidth]{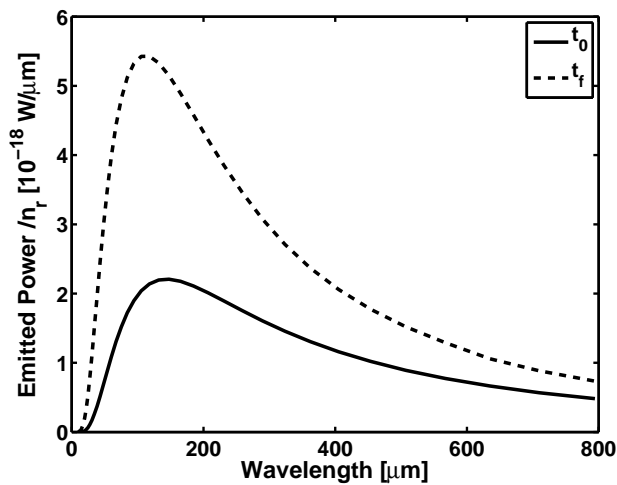}
\end{center} \vspace{-10mm}
\caption{\small \it \noindent Average synchrotron spectrum emitted by the initial runaway population ($t_\mathrm{0}$, solid line) and after interaction with the EXEL wave ($t_\mathrm{f}$, dashed), for the JET-like reference scenario parameters.}
	\label{fig:sync_JET}
\end{figure}

The result of the synchrotron spectrum calculation for the reference
JET-like scenario is shown in Fig.~\ref{fig:sync_JET}. As an initial
distribution, Eq.~\eqref{eq:initial} was used with $R=3$~m. The
distribution affected by the interaction was taken at $604\;\rm \mu s$
after the first destabilization of the most unstable wave (this is the 
distribution function plotted in Fig.~\ref{fig:changing_JET}). The
runaway region in momentum space ($S_{RE}$) was defined by $p_\|\in
[12,31]$ and $p_\perp \in [0,3]$
(cf. Fig.~\ref{fig:changing_JET}). Due to the strong energy dependence
of the synchrotron emission, and the exponential fall-off of the
distribution with increasing $p_\perp$, contributions from particles
with lower parallel momentum or larger perpendicular momentum were
negligible. The maximal parallel momentum was determined by the
cut-off of the distribution function
\eqref{eq:sol} at $p_{\mathrm{max}}+\sigma_p$.

The synchrotron spectra in Fig.~\ref{fig:sync_JET} show that in $604\;\rm \mu s$, the particle-wave interaction  has a significant effect on the synchrotron spectrum, with the peak emission increasing by roughly a factor 2.5. The wavelength of peak emission is also shifted slightly towards shorter wavelengths. We emphasize that Fig.~\ref{fig:sync_JET} shows the average emission per runaway, meaning that if the number of runaways can be considered fixed, the effect of the interaction with the EXEL wave is a significant increase of the total synchrotron emission.  

As discussed in Ref.~\cite{adamsyrup}, however, there are several other factors that have a similar effect on the synchrotron spectrum. The spectrum is highly dependent on the properties of the runaway distribution, and is thus sensitive to plasma parameters such as temperature, density, impurity content and electric and magnetic fields. The effect of the EXEL wave interaction could only be discriminated by the characteristic sudden increase of the emission on the $100-1000\;\rm \mu s$ time scale.

The EXEL wave interaction is likely to produce an even larger effect
than that shown in Figures \ref{fig:changing_JET} and
\ref{fig:sync_JET}, however, since as the assumption of beam-like
distribution used in our modeling breaks down, we can not simulate the
later stages of the interaction. Pitch-angle scattering might also
increase radial transport and eventually lead to the mitigation of the
runaway beam, but at an earlier stage our model predicts a burst of
synchrotron radiation unaccompanied by macroscopic MHD activity.

For the distribution used in Fig.~\ref{fig:sync_JET}, the wavelength
region of strong emission is in the far-infrared and sub-millimeter
regions of the spectrum, implying that detection of the effect of the EXEL wave
instability by means of synchrotron radiation would require an
infrared camera sensitive to this wavelength range. The reason for the long wavelength emission is the cut-off of the runaway distribution (Eq.~\ref{eq:initial}) at a particle energy of roughly 15~MeV. A runaway
electron distribution extending to higher maximum energy would allow
detection by ordinary near-infrared (or even visible light) cameras,  but realistic modeling of the evolution of the distribution function in a disruption is out of
the scope of this paper.

\section{Conclusions}
\label{sec:conclusions}

Runaway electrons pose a significant threat to tokamaks. This is
especially true for ITER, where the runaway current might be as high
as $12 \;\rm MA$ in disruptions, with the electron energy spectrum
extending up to several tens of MeV \cite{hollmann}.  In this paper
the quasi-linear resonant interaction of the runaway population and
the high-frequency obliquely propagating extraordinary-electron (EXEL)
wave, which leads to rapid pitch-angle scattering of the resonant
runaways, was studied. The scattering occurs when the runaway density
reaches a certain critical density of about $10^{14} - 10^{17} \;\rm
m^{-3}$, depending on the plasma parameters. As soon as the EXEL wave
is destabilized, it leads to a pitch-angle scattering of resonant
electrons through quasi-linear diffusion in the velocity space on the
$100-1000\;\rm \mu s$ time scale. In our simulations, the spectral energy of the
destabilized EXEL wave did not exceed $~10^{-11} \;\rm J$ in any of the scenarios considered, implying that direct experimental detection of the wave is
likely to be difficult.

As the resonant interaction with the EXEL wave mainly affects the high energy runaways, which are the electrons characterizing the synchrotron radiation emitted by the whole population, the interaction causes a significant change in the synchrotron spectrum. The interaction with the EXEL wave was shown to produce a burst of synchrotron radiation accompanied by a simultaneous shift of the spectrum towards shorter wavelengths, which might offer a possibility to detect the impact of the quasi-linear interaction in experiments.

By looking at a wide range of plasma parameters, we concluded that the
characterizing quantities of the interaction (resonant runaway
momentum, wave energy, critical runaway density, etc.) have a weak
dependency on plasma parameters. We can therefore extend our conclusions to
an ITER-like scenario. The intensity of the interaction, and the
resulting change in the synchrotron spectrum, are expected to be
qualitatively similar and of the same order of magnitude as for the
investigated JET-like reference scenario. The only major difference
in the ITER case is a slightly higher stability threshold (which could still easily be reached) due to he stronger
magnetic field.

Our analysis shows that the EXEL wave is destabilized considerably more
easily, as compared to the previously studied whistler wave
\cite{fuloppop2006,pokolppcf2008}. A possibility for experimental confirmation of the results presented in this paper is offered through the predicted bursts in
the far-infrared synchrotron emission. Our results also provide a basis for
further theoretical work making use of more realistic kinetic
simulations with advanced Fokker-Planck solvers, such as the LUKE code \cite{LUKE}.

\section*{Acknowledgments}
The authors are grateful to J.~Decker, Y.~Peysson and G.~Papp for fruitful
discussions. This project has received funding from the European
Union’s Horizon 2020 research and innovation programme under grant
agreement number 633053. The views and opinions expressed herein do
not necessarily reflect those of the European Commission

\end{document}